\documentclass[twocolumn,showpacs,preprintnumbers,amsmath,amssymb]{revtex4}

\usepackage{epsfig}

\newtheorem{theorem}{Theorem}
\newtheorem{lemma}{Lemma}

\begin{document}


\title{The Clifford group, stabilizer states, and linear and
quadratic operations over GF(2). }

\author{Jeroen Dehaene}
\email{Jeroen.Dehaene@esat.kuleuven.ac.be}
\affiliation{Katholieke Universiteit Leuven, ESAT-SCD, Belgium}
\author{Bart De Moor}
\affiliation{Katholieke Universiteit Leuven, ESAT-SCD, Belgium}
\date{\today}

\begin{abstract}
  We describe stabilizer states and Clifford group operations using
  linear operations and quadratic forms over binary vector spaces. We
  show how the $n$-qubit Clifford group is isomorphic to a group with
  an operation that is defined in terms of a $(2n+1)\times
  (2n+1)$ binary matrix product and binary quadratic forms. As an
  application we give two schemes to efficiently decompose Clifford
  group operations into one and two-qubit operations. We also show how
  the coefficients of stabilizer states and Clifford group operations
  in a standard basis expansion can be described by binary quadratic
  forms. Our results are useful for quantum error correction,
  entanglement distillation and possibly quantum computing.
\end{abstract}


\maketitle

\section{Introduction}

Stabilizer states and Clifford group operations play a central role in
quantum error correction, quantum computing, and entanglement
distillation. A stabilizer state is a state of an $n$-qubit system
that is a simultaneous eigenvector of a commutative subgroup of the
Pauli group. The latter consists of all tensor products of $n$
single-qubit Pauli operations. The Clifford group is the group of
unitary operations that map the Pauli group to itself under
conjugation. In quantum error correction these concepts play a
central role in the theory of stabilizer codes
\cite{Got:97}. Although a quantum computer working with only
stabilizer states and Clifford group operations is not powerful enough
to disallow efficient simulation on a classical computer \cite{ChN:00,
Got:98}, it is not unlikely that possible new quantum algorithms will
exploit the rich structure of this group. In \cite{DVD:03}, we also
showed the relevance of a quotient group of the Clifford group in
mixed state entanglement distillation.

In this paper, we link stabilizer states and Clifford operations with
binary linear algebra and binary quadratic forms (over GF(2)). The
connection between multiplication of Pauli group elements and binary
addition is well known as is the connection between commutability of
Pauli group operations and a binary symplectic inner product
\cite{Got:97}. In \cite{DVD:03} we extended this connection to a
link between a quotient group of the Clifford group and binary
symplectic matrices (there termed $P$ orthogonal). In this paper we
give a binary characterization of the full Clifford group, by adding
quadratic forms to the symplectic operations. In addition we show how
the coefficients, with respect to a standard basis, of both stabilizer
states and Clifford operations can also be described using binary
quadratic forms. Our results also lead to efficient ways for decomposing
Clifford group operations in a product of 2-qubit operations.

\section{Clifford group operations and binary linear and quadratic
operations}
\label{secbin}
In this section, we show how the Clifford group is isomorphic to a
group that can be entirely described in terms of binary linear algebra,
by means of symplectic linear operations and quadratic forms.

We use the following notation for Pauli matrices.
\[
\begin{array}{lrrl}
 \sigma_{00} & = \tau_{00} & =\sigma_0 & =
             \left[\begin{array}{rr} 1 & 0\\
                                     0 & 1
             \end{array}\right],\\
 \sigma_{01} & = \tau_{01} & =\sigma_x & =
             \left[\begin{array}{rr} 0 & 1\\
                                     1 & 0
             \end{array}\right],\\
 \sigma_{10} & = \tau_{10} & =\sigma_z & =
             \left[\begin{array}{rr} 1 & 0\\
                                     0 & -1
             \end{array}\right],\\
 \sigma_{11} &             & =\sigma_y & =
             \left[\begin{array}{rr} 0 & -i\\
                                    i & 0
             \end{array}\right],\\
             & \tau_{11}   & =i\sigma_y & =
             \left[\begin{array}{rr} 0 & 1\\
                                    -1 & 0
             \end{array}\right].
\end{array}
\]

We also use vector indices to indicate tensor products of Pauli
matrices. If $v,w\in\mathbb{Z}_2^{n}$ and
$a=\left[\begin{array}{c}v\\w\end{array}\right]\in\mathbb{Z}_2^{2n}$,
then we denote
\begin{equation}
\label{eqsigtau}
\begin{array}{ll}
 \sigma_a &=\sigma_{v_1w_1}\otimes\ldots\otimes\sigma_{v_nw_n},\\
 \tau_a &=\tau_{v_1w_1}\otimes\ldots\otimes\tau_{v_nw_n}
\end{array}
\end{equation}

If we define the Pauli group to contain all tensor products of Pauli
matrices with an additional complex phase in $\{1,i,-1,-i\}$, an
arbitrary Pauli group element can be represented as $i^\delta
(-1)^\epsilon \tau_u$, where $\delta,\epsilon\in\mathbb{Z}_2$ and
$u\in\mathbb{Z}_2^{2n}$. The separation of $\delta$ and $\epsilon$,
rather than having $i^\gamma$ with $\gamma\in\mathbb{Z}_4$, is
deliberate and will simplify formulas below. Throughout this paper
exponents of $i$ will always be binary. As a result
$i^{\delta_1}i^{\delta_2}=i^{\delta_1+\delta_2}(-1)^{\delta_1\delta_2}$.
Multiplication of two Pauli group elements can now be translated into
binary terms in the following way:
\begin{lemma}
\label{lemtautau}
If $a_1,a_2\in\mathbb{Z}_2^{2n}$,
$\delta_1,\delta_2,\epsilon_1,\epsilon_2\in\mathbb{Z}_2$ and $\tau$ is
defined as in Eq.~(\ref{eqsigtau}), then
\[
\begin{array}{l}
 i^{\delta_1} (-1)^{\epsilon_1} \tau_{a_1} i^{\delta_2}
 (-1)^{\epsilon_2} \tau_{a_2} =i^{\delta_{12}}(-1)^{\epsilon_{12}}
 \tau_{a_{12}}\\
 \mbox{with} \begin{array}[t]{ll}
   \delta_{12} & = \delta_1+\delta_2\\
   \epsilon_{12} & =\epsilon_1+\epsilon_2+\delta_1\delta_2 + a_2^TUa_1\\
   a_{12} & = a_1+a_2,\\
   U & =\left[\begin{array}{rr}
     0_n & I_n\\
     0_n & 0_n
     \end{array}\right],
 \end{array}
\end{array}
\]
where multiplication and addition of binary variables is modulo 2.
\end{lemma}

These formulas can easily be verified for $n=1$ and then
generalized for $n>1$. The term $a_2^TUa_1$ ``counts'' (modulo 2) the
number of positions $k$ where ${w_1}_k=1$ and ${v_2}_k=1$ (with
$a_1=\left[\begin{array}{c}v_1\\w_1\end{array}\right]$ and
$a_2=\left[\begin{array}{c}v_2\\w_2\end{array}\right]$), as only these
positions get a minus sign in the following derivation:
\[
\begin{array}{rl}
\tau_{{v_1}_k {w_1}_k}\tau_{{v_2}_k {w_2}_k} &
=\sigma_z^{{v_1}_k}\sigma_x^{{w_1}_k}\sigma_z^{{v_2}_k}\sigma_x^{{w_2}_k}\\
&=(-1)^{{w_1}_k{v_2}_k}\sigma_z^{{v_1}_k+{v_2}_k}\sigma_x^{{w_1}_k+{w_2}_k}\\
&=(-1)^{{w_1}_k{v_2}_k}\tau_{{v_1}_k+{v_2}_k,{w_1}_k+{w_2}_k}.
\end{array}
\]

A Clifford group operation $Q$, by definition, maps the Pauli group to
itself under conjugation:
\[
  Q \tau_a Q^\dag = i^\delta (-1)^\epsilon \tau_b
\]
for some $\delta$,$\epsilon$,$b$, function of $a$.

Because
$Q\tau_{a_1}\tau_{a_2}Q^\dag=(Q\tau_{a_1}Q^\dag)(Q\tau_{a_2}Q^\dag)$,
it is 
sufficient to know the image of a generating set of the Pauli group to
know the image of all Pauli group elements and define $Q$ (up to an
overall phase). In binary terms it is sufficient to know the image of
$\tau_{b_k},~k=1,\ldots,n$ where $b_k,~k=1,\ldots,n$ form a basis of
$\mathbb{Z}_2^{2n}$.

For this purpose it is possible to work with Hermitian Pauli group elements
only as the image of a Hermitian matrix under $X\rightarrow QXQ^\dag$
will again be Hermitian (and the images of Hermitian Pauli group elements are
sufficient do derive the images of non Hermitian ones).
In our binary language Hermitian Pauli group elements are described as
\[
  i^{a^TUa}(-1)^\epsilon\tau_a
\]
as $a^TUa$ counts (modulo 2) the number of $\tau_{11}$ in the tensor
product $\tau_a$. For $\tau_{11}$ is the only non-Hermitian (actually
skew Hermitian) of the four $\tau$ matrices and multiplication with
$i$ makes it Hermitian.

Now we take the standard basis of $\mathbb{Z}_2^{2n}$
$e_k,~k=1,\ldots,n$ where $e_k$ is the $k$-th column of $I_{2n}$, and
consider the generating set of Hermitian operators $\tau_{e_k}$. These
correspond to single-qubit operations $\sigma_z$ and $\sigma_x$. We
denote their images under $X\rightarrow QXQ^\dag$ by
$i^{d_k}(-1)^{h_k} \tau_{c_k}$ and assemble the vectors $c_k$ in a
matrix $C$ (with columns $c_k$) and the scalars $d_k$ and $h_k$ in the
vectors $d$ and $h$. As the images are Hermitian, $d_k=c_k^TUc_k$ or
$d=\mbox{diag}(C^TUC)$ (with $\mbox{diag}(X)$ the vector with the
diagonal elements of $X$).

Now, given $C$, $d$ and $h$, defining the Clifford operation $Q$, the
image $i^{\delta_2}(-1)^{\epsilon_2}\tau_{b_2}$ of
$i^{\delta_1}(-1)^{\epsilon_1}\tau_{b_1}$ under $X\rightarrow
QXQ^\dag$ can be found by mutliplying those operators
$i^{d_k}(-1)^{h_k} \tau_{c_k}$ for which ${b_1}_k=1$. By repeated
application of Lemma~\ref{lemtautau}, this yields
\[
 \begin{array}{ll}
   b_2 &=C b_1\\
   \delta_2 &= \delta_1 + d^T b_1\\
   \epsilon_2 &= \epsilon_1 + h^T b_1 +
   b_1^T(\mbox{lows}(C^TUC+dd^T)b_1 + \delta_1 d^T b_1
 \end{array}
\]
where $\mbox{lows}(X)$ is the strictly lower triangular part of $X$.
These formulas can be simplified by introducing the following notation
\[
 \begin{array}{llll}
   \bar C &=\left[\begin{array}{ll}
              C   & 0\\
              d^T & 1
            \end{array}\right]\\
   \bar U &=\left[\begin{array}{rr}
              U   & 0\\
              0   & 1
            \end{array}\right]\\
   \bar h   &=\left[\begin{array}{r} h\\ 0 \end{array}\right]\\
   \bar b_1 &=\left[\begin{array}{r} b_1\\ \delta_1 \end{array}\right] &
   \bar b_2 &=\left[\begin{array}{r} b_2\\ \delta_2 \end{array}\right]\\
   \tau_{\bar b_1} &= i^{\delta_1} \tau_{b_1} &
   \tau_{\bar b_2} &= i^{\delta_2} \tau_{b_2}
 \end{array}
\]
We then get the following theorem
\begin{theorem}
\label{theocb}
  Given $\bar C$ and $\bar h$, defining the Clifford operation $Q$ as
  above, the image under $X\rightarrow QXQ^\dag$ of 
  $(-1)^{\epsilon_1}\tau_{\bar b_1}$ is
  $(-1)^{\epsilon_2}\tau_{\bar b_2}$ with
\[
 \begin{array}{ll}
   \bar b_2 &= \bar C \bar b_1\\
   \epsilon_2 &= \epsilon_1 + \bar h^T \bar b_1 + \bar b_1 \mbox{lows}(\bar C^T
   \bar U \bar C) \bar b_1
 \end{array}
\]
\end{theorem}

With this theorem we can also compose two Clifford operations using
the binary language. To this end we have to find the images under the
second operation of the images under the first operation of the
standard basis vectors. This can be done using Theorem~\ref{theocb}:

\begin{theorem}
\label{theoQ21}
Given $\bar C_1$, $\bar h_1$, $\bar C_2$ and $\bar h_2$, defining two Clifford
operations $Q_1$ and $Q_2$ as above, the product $Q_{21}=Q_2 Q_1$ is
represented by $\bar C_{21}$ and $\bar h_{21}$ given by
\[
 \begin{array}{ll}
   \bar C_{21}&=\bar C_2 \bar C_1\\
   \bar h_{21}&=\bar h_1 + \bar C_1^T \bar h_2 + \mbox{diag}(\bar C_1^T
   \mbox{lows}(\bar C_2^T \bar U \bar C_2)\bar C_1)
 \end{array}
\]

\end{theorem}

The next question is which $\bar C$ and $\bar h$ or $C$, $d$ and $h$
can represent a Clifford operation. The answer is that $C$ has to be a
symplectic matrix (and $d$ has to be equal to $\mbox{diag}(C^TUC)$ as
above). If we define $P$ to be $U+U^T$, we call a matrix symplectic if
$C^TPC=P$. One way to see that $C$ has to be symplectic is through the
connection of the symplectic inner product $b^TPa$ with commutability
of Pauli group elements:
\[
 \tau_a\tau_b=(-1)^{b^TPa}\tau_b\tau_a
\]
Since the map $X\rightarrow QXQ^\dag$ preserves commutability, $a$ and
$b$ have to represent commutable Pauli group elements ($b^TPa=0$) if
and only if $Ca$ and $Cb$ represent commutable elements
($b^TC^TPCa=0$). This implies that $C$ has to be symplectic.

That symplecticity is also sufficient was first implied by Theorem~1
of \cite{DVD:03} (almost, as this result was set in the context of
entanglement distillation where the signs $\epsilon$ play no
significant role). The idea is to give a constructive way of realizing
the Clifford operation $Q$ given by $\bar C$ and $\bar h$.  This can
be done using only one and two-qubit operations, which makes the result also
of practical use. In Sec.~\ref{sec2q} we give two such decompositions
that are more transparent than the results of \cite{DVD:03}.

First, to conclude this section, we complete the binary group picture
by a formula for the inverse of a Clifford group element, given in
binary terms. 

\begin{theorem}
\label{theoinv}
Given $\bar C_1$ and $\bar h_1$, defining a Clifford operation $Q_1$ as above,
the inverse $Q_2=Q_1^{-1}$ is represented by
\[
 \begin{array}{ll}
  \bar C_2&=\bar C_1^{-1}=
     \left[\begin{array}{ll}
       C_1^{-1} & 0\\
       d^T C^{-1} & 1
     \end{array}\right] =
     \left[\begin{array}{ll}
       PC_1^TP & 0\\
       d_1^TPC_1^TP & 1
     \end{array}\right] \\
  \bar h_2&=\bar C^{-T} \bar h +
       \mbox{diag}(\bar C^{-T}\mbox{lows}(\bar C^T\bar U\bar C)\bar C^{-1})
 \end{array}
\]
\end{theorem}

These formulas can be verified using Theorem~\ref{theoQ21}.  Finally
note that since the Clifford operations form a group and the matrices
$\bar C$ are simply multiplied when composing Clifford group
operations, the matrices $\bar C$ with $C$ symplectic and
$d=\mbox{diag}(C^TUC)$ must form a group of $(2n+1)\times(2n+1)$
matrices that is isomorphic to the symplectic group of $2n\times 2n$
matrices.  This can be easily verified by showing that
\[
 \mbox{diag}(C_1^TC_2^TUC_2C_1)=C_1^T\mbox{diag}(C_2^TUC_2)+\mbox{diag}(C_1^TUC_1)
\]
This follows from the fact that $C^TUC+U$ is symmetric when $C^TPC=P$
and $x^TSx=x^T\mbox{diag}(S)$ when $S$ is symmetric. In a similar way
it can be proven that
$\mbox{diag}(C^{-T}UC^{-1})=C^{-T}\mbox{diag}(C^TUC)$.

\section{Special Clifford operations in the binary picture}
\label{secspec}
In this section we consider a selected set of Clifford group operations
and their representation in the binary picture of Sec.~\ref{secbin}.

First, we consider the Pauli group operations $Q=\tau_a$ as Clifford
operations. Note that a global phase cannot be represented as it does
not affect the action $X\rightarrow QXQ^\dag$. To construct $C$ and
$h$ we have to consider the images of the operators $\tau_{e_k}$
representing one-qubit operations $\sigma_x$ and $\sigma_z$. One can
easily verify that $\tau_a$ is represented by
\begin{equation}
\label{eqsc1}
 \begin{array}{ll}
  C&=I_{2n}\\
  h&=Pa
 \end{array}
\end{equation}

Second, note that Clifford operations acting on a subset $\alpha
\subset \{1,\ldots,n\}$ consist of a symplectic matrix on the rows and
columns with indices in $\alpha\cup(\alpha+n)$, embedded in an identity matrix
(that is, with ones on positions $C_{k,k}=1$,
$k\not\in\alpha\cup(\alpha+n)$ and $C_{k,l}=0$ if $k\neq l$ and $k$ or
$l$ $\not\in\alpha\cup(\alpha+n)$.) Also $h_k=0$ if
$k\not\in\alpha\cup(\alpha+n)$.

Third, qubit permutations, are represented by
\[
 \begin{array}{ll}
 C&=\left[\begin{array}{ll}
     \Pi & 0\\
     0 & \Pi
   \end{array}\right]\\
 h&=0
 \end{array}
\]
where $\Pi$ is a permutation matrix.

Fourth, the conditional not or CNOT operation on two qubits is
represented by
\[
 \begin{array}{ll}
   C&= \left[\begin{array}{llll}
         1 & 1 & 0 & 0\\
         0 & 1 & 0 & 0\\
         0 & 0 & 1 & 0\\
         0 & 0 & 1 & 1\\
       \end{array}\right]\\
   h&=0
 \end{array}
\]

Fifth, by composing qubit permutations and CNOT operations on selected
qubits any linear transformation of the index space
$|x\rangle\rightarrow|Rx\rangle$ can be realized, where
$x\in\mathbb{Z}_2^n$ labels the standard basis states
$|x\rangle=|x_1\rangle\otimes\ldots\otimes|x_n\rangle$ and
$R\in\mathbb{Z}_2^{n\times n}$ is an invertible matrix (modulo
2). This operation is represented in the symplectic picture by
\begin{equation}
\label{eqsc5}
 \begin{array}{ll}
 C&=\left[\begin{array}{ll}
     R^{-T} & 0\\
     0      & R
   \end{array}\right]\\
 h&=0
 \end{array}
\end{equation}

The qubit permutations and CNOT operation discussed above are special
cases of such operations as qubit permutations can be represented as
$|x\rangle\rightarrow |\Pi x\rangle$ and the CNOT operation as
$|x\rangle\rightarrow
|\left[\begin{array}{ll}
   1 & 0\\
   1 & 1
 \end{array}\right]x\rangle$.

Decomposing a general linear transformation $R$ into CNOTS and
qubit permutations can be done by Gauss elimination (a well known
technique for the solution of systems of linear equations). In this
process $R$ is operated on on the left by CNOTS and qubit permutations to
be gradually transformed in an identity matrix. The process operates
on $R$, column by column, first moving a nonzero element into the
diagonal position by a qubit permutation, then zeroing the rest of the
column by CNOTS. The inverses of the applied operations yield a
decomposition of $R$.

Sixth, we consider Hadamard operations. The Hadamard operation on a
single qubit
$Q=H=\frac{1}{\sqrt{2}}\left[\begin{array}{rr}1&1\\1&-1\end{array}\right]$ 
is represented by
$C=\left[\begin{array}{ll}0 & 1\\1 & 0\end{array}\right]$ and $h=0$.
A Hadamard operation on a selected set of qubits is represented by the
embedding of such matrices in an identity matrix as explained above.
As a special case we mention the Hadamard operation on all qubits,
which is represented by $C=P$ and $h=0$.

Seventh, we consider operations
$e^{i(\pi/4)\tau_{\bar a}}=\frac{1}{\sqrt{2}}(I+i\tau_{\bar a})$ where
  $a\in\mathbb{Z}_2^{2n}$,
  $\bar a=\left[\begin{array}{c}a\\a^TUa\end{array}\right]$, and
  $\tau_{\bar a}=i^{a^TUa}\tau_a$.
 These operations are represented by
\begin{equation}
\label{eqsc7}
 \begin{array}{ll}
   C&=I+aa^TP\\
   h&=C^TUa
 \end{array}
\end{equation}
This is proved in the Appendix.

Finally, we mention that real Clifford operations have $d=0$.

\section{Decompositions of Clifford operations in one and two-qubit operations}
\label{sec2q}

In this section we write general Clifford group operations as products
of one and two-qubit operations using the binary picture. This does not only
complete the results of Sec.~\ref{secbin}, showing that every
symplectic $C$ and arbitrary $h$ represent a Clifford operation. It is
also of practical use for quantum computing applications as well as
entanglement distillation applications since two-qubit operations can
be realized relatively easily and the number of two-qubit operations
needed is ``only'' quadratical in the number of qubits. We give two
different schemes.

First, for both schemes, we observe that the main problem is realizing
$C$, not $h$.  For once a Clifford operation represented by $C$ and
$h'$ is realized, we can realize $h$ by doing an extra operation
$Q=\tau_{CP(h+h')}$ on the left or $Q=\tau_{P(h+h')}$ on the right. This
can be proved by using Eq.~(\ref{eqsc1}) and Theorem~\ref{theoQ21}.

The first scheme realizes $C$ by two-qubit operations, acting on qubit
$k$ and $l$, of the type $e^{i(\pi/4)\tau_{\bar a}}$ with symplectic
matrices $(I+aa^TP)$ where $a$ can be nonzero (i.e. one) only at
positions $k,l,n+k$ and $n+l$. The scheme works by reducing a given
symplectic matrix $C$ to the identity matrix by operating on the left
with two-qubit operations. The product of the inverses of these
two-qubit matrices is then equal to $C$. The reduction to the identity
matrix is done by working on two columns $m$ and $n+m$ at a time, for
$m=1,\ldots,n$. First columns $1$ and $n+1$ are reduced to columns $1$
and $n+1$ of the identity matrix. Because through all the operations
$C$ remains symplectic, one can show that as a result also rows $1$
and $n+1$ are reduced to rows $1$ and $n+1$ of the identity matrix.
Then one can repeat the same process on the submatrix of $C$ obtained
by dropping rows and columns $1$ and $n+1$, until the whole matrix is
reduced to the identity matrix.

Let $\alpha=\{1,1+n\}$ and $\beta=\{l,l+n\}$. The first step in
reducing columns $1$ and $n+1$ of $C$ to the corresponding columns of
the identity matrix is a qubit permutation, exchanging qubit $1$ with
some qubit $k$ to make $C_{\alpha,\alpha}$ invertible. This can be
done for if all $C_{\beta,\alpha}$ would be rank deficient, we would
have $c_1^TPc_{n+1}=0$ which is in conflict with the symplecticity of
$C$.  (Note that a $2\times 2$-matrix is invertible if and only if it
is symplectic). Next, we perform two-qubit operations
$e^{i(\pi/4)\tau_{\bar a}}$ on qubits $1$ and $l$ with
$a_{\alpha}=c_{\alpha,n+1}$ and $a_{\beta}=c_{\beta,1}$, for
$l=2,\ldots,n$. Such an operation changes $C$ through multiplication
with $I+aa^TP$. For the first column this means that $c_1$ is replaced
by $c_1+a$, as $a^TPc_1= c_{\alpha,n+1}^TP_2c_{\alpha,1}+
c_{\beta,1}^TP_2c_{\beta,1}= 1+0=1$, where
$P_2=\left[\begin{array}{cc}0&1\\1&0\end{array}\right]$. This way
$c_{\beta,1}$ is reduced to $0$. $C_{\alpha,\alpha}$ is changed at
every step but remains invertible (and symplectic). Note that through
these operations also the other columns of $C$ are changed. After the
first column has been zeroed on all positions except $\alpha$, we
tackle column $n+1$ with operations $e^{i(\pi/4)\tau_{\bar a}}$ on
qubits $1$ and $l$ with $a_{\alpha}=c_{\alpha,1}$ and
$a_{\beta}=c_{\beta,n+1}$, $l=2,\ldots,n$. These operations have no
effect on $c_1$ because $a^TPc_1=c_{\alpha,1}^TP_2c_{\alpha,1}+0=0$,
and reduce $c_{\beta,n+1}$ to $0$ in the same way as was done for the
first column. After these operations we are left with $c_1$ and
$c_{n+1}$ all $0$ except for $C_{\alpha,\alpha}$ which equals an
invertible matrix. This matrix can be transformed into an identity
matrix by a one-qubit symplectic operation on qubit $1$. One-qubit
Clifford operations can be easily made by one-qubit operations of type
$e^{i(\pi/4)\tau_{\bar a}}$.

An advantage of this scheme is that it is efficient if only some
columns of $C$ (or rows, as one can also work on the right) are
specified while the other columns do not matter. This is the case in
the entanglement distillation protocols of \cite{DVD:03}.

The second scheme also takes a number of steps that is quadratical in
$n$. It is based on the following theorem, which will also be of
importance in Sec.~\ref{secdesc}, and for which we give
a constructive proof.

\begin{theorem}
\label{theosympdec}
If $C\in\mathbb{Z}_2^{2n\times 2n}$ is a symplectic matrix
($C^TPC=P$), it can be decomposed as
\begin{eqnarray}
 C&=&\left[\begin{array}{cc}
     T_1^{-T} & 0\\
     0        & T_1
    \end{array}\right]\times\label{eqsd1}\\
   && \left[\begin{array}{llll}
       I_{n-r} & V_1  & Z_3+V_1V_2^T   & V_2+V_1Z_2\\
       0       & Z_1  & V_1^T+Z_1V_2^T & I_r+Z_1Z_2\\
       0       & 0    & I_{n-r}        & 0\\
       0       & I_r  & V_2^T          & Z_2
    \end{array}\right]
    \left[\begin{array}{cc}
     T_2^{-T} & 0\\
     0        & T_2
    \end{array}\right]\nonumber\\
  &=&\left[\begin{array}{cc}
     T_1^{-T} & 0\\
     0        & T_1
    \end{array}\right]\times\label{eqsd2}\\
  &&\left[\begin{array}{llll}
      I_{n-r} & 0   & Z_3     & V_1\\
      0       & I_r & V_1^T   & Z_1\\
      0       & 0   & I_{n-r} & 0\\
      0       & 0   & 0       & I_r
    \end{array}\right]
    \left[\begin{array}{llll}
      I_{n-r} & 0   & 0       & 0\\
      0       & 0   & 0       & I_r\\
      0       & 0   & I_{n-r} & 0\\
      0       & I_r & 0       & 0
    \end{array}\right]\times\nonumber\\
  &&\left[\begin{array}{llll}
      I_{n-r} & 0   & 0       & V_2\\
      0       & I_r & V_2^T   & Z_2\\
      0       & 0   & I_{n-r} & 0\\
      0       & 0   & 0       & I_r
    \end{array}\right]
     \left[\begin{array}{cc}
     T_2^{-T} & 0\\
     0        & T_2
    \end{array}\right]\nonumber
\end{eqnarray}
where $T_1$ and $T_2$ are invertible $n\times n$ matrices,
$Z_1$ and $Z_2$ are symmetric $r\times r$ matrices, $Z_3$ is a
symmetric $(n-r)\times(n-r)$ matrix, $V_1$ and $V_2$ are
$(n-r)\times r$ matrices and the zero blocks have appropriate
dimensions.
\end{theorem}

{\bf Proof:}
To prove this theorem We consider $C$ as a block matrix
$C=\left[\begin{array}{cc} E' & F'\\
                           G' & H'\\
         \end{array}\right]$.

Then, we find invertible $R_1$ and $R_2$ in $\mathbb{Z}_2^{n\times n}$ such
         that $R_1^{-1}G'R_2=
         \left[\begin{array}{cc} 0 & 0\\
                                 0 & I_{r}
         \end{array}\right],$ where $r$ is the rank of $G'$. This
is a standard linear algebra technique and can be realized (for
example) by (1) setting the first $n-r$ columns of $R_2$
equal to a basis of the kernel of $G'$, (2) choosing the other
columns of $R_2$ as to make it invertible, (3) setting the last $r$
columns of $R_1$ equal to the last $r$ columns of $R_2$ multiplied on the
left by $G'$ (This yields a basis of the range of $G'$), and (4)
choosing the other columns of $R_1$ as to make it invertible. By
construction, this implies
  $G'R_2=R_1\left[\begin{array}{cc} 0 & 0\\ 
                               0 & I_{r}
       \end{array}\right]$.

Now we set
\begin{equation}
\label{eqefh0}
\begin{array}{l}
  \left[\begin{array}{cc}
    R_1^T    & 0\\
    0        & R_1^{-1}
  \end{array}\right]
  C
  \left[\begin{array}{cc}
    R_2 & 0\\
    0   & R_2^{-T}
  \end{array}\right]=\\

  ~\left[\begin{array}{llll}
      E_{11} & E_{12} & F_{11} & F_{12}\\
      E_{21} & E_{22} & F_{21} & F_{22}\\
      0      & 0      & H_{11} & H_{12}\\
      0      & I_r    & H_{21} & H_{12}
   \end{array}\right]
\end{array}
\end{equation}
Because the three matrices in the left-hand side of Eq.~(\ref{eqefh0})
are symplectic, so is the right-hand side. This leads to the following
relations between its submatrices:
\begin{eqnarray}
 E_{21}^T=0\label{eqefh1}\\
 E_{11}^TH_{11}+E_{21}^TH_{21}=I\label{eqefh2}\\
 E_{11}^TH_{12}+E_{21}^TH_{22}=0\label{eqefh3}\\
 E_{22}^T+E_{22}=0\label{eqefh4}\\
 E_{12}^TH_{11}+E_{22}^TH_{21}+F_{21}=0\label{eqefh5}\\
 E_{12}^TH_{12}+E_{22}^TH_{22}+F_{22}=I\label{eqefh6}\\
 F_{11}^TH_{11}+F_{21}^TH_{21}+H_{11}^TF_{11}+H_{21}^TF_{21}=0\label{eqefh7}\\
 F_{11}^TH_{12}+F_{21}^TH_{22}+H_{11}^TF_{12}+H_{21}^TF_{21}=0\label{eqefh8}\\
 F_{12}^TH_{12}+F_{22}^TH_{22}+H_{12}^TF_{12}+H_{22}^TF_{22}=0\label{eqefh9}
\end{eqnarray}

With Eq.~(\ref{eqefh1}) and Eq.~(\ref{eqefh2}) we find $H_{11}$=$E_{11}^{-T}$.
Now, if we replace $R_2$ by
  $R_2 \left[\begin{array}{ll}
      E_{11}^{-1} & 0\\
      0      & I_r
   \end{array}\right]$,
both $H_{11}$ and $E_{11}$ are replaced by $I_{n-r}$. We will
assume that this choice of $R_2$ was taken from the start. Then, from
Eq.~(\ref{eqefh1}) and Eq.~(\ref{eqefh3}) we find $H_{12}=0$. From
Eq.~(\ref{eqefh4}) we learn that $E_{22}$ is symmetric. From
Eq.~(\ref{eqefh5}) and Eq.~(\ref{eqefh6}) we find
$F_{21}=E_{12}^T+E_{22}^TH_{21}$ and
$F_{22}=I+E_{22}H_{22}$. Substituting these equations in
Eqs.~(\ref{eqefh7}),(\ref{eqefh8}) and~(\ref{eqefh9}), we find that
$F_{11}+H_{21}^TE_{12}^T$ is symmetric, $F_{12}=H_{21}^T+E_{12}H_{22}$,
and $H_{22}$ is symmetric. Setting $T_1=R_1$, $T_2=R_2^T$ (with $R_2$
chosen as to make $E_{11}=H_{11}=I$), $V_1=E_{12}$, $V_2=H_{21}^T$,
$Z_1=E_{22}$, $Z_2=H_{22}$ and $Z_3=F_{11}+V_1V_2^T$, we obtain 
Eq.~(\ref{eqsd1}). Note that $Z_3$ is symmetric because
$F_{11}+V_2V_1^T$ and $V_2V_1^T+V_1V_2^T$ are symmetric. Finally
Eq.~\ref{eqsd2} can be easily verified. This completes the
proof. \hfill$\square$

To find a decomposition of $C$ in one and two-qubit operations
we concentrate on the five matrices in the right-hand side of
Eq.~(\ref{eqsd2}), all of which are symplectic. Clearly the first and
last matrix are linear index space transformations as discussed in
Sec.~\ref{secspec}. These can be decomposed into CNOTs and qubit
permutations. The middle matrix corresponds to Hadamard operations on the
last $r$ qubits. We will now show that the second and fourth matrix
can be realized by one and two-qubit operations of the type
$e^{i(\pi/4)\tau_{\bar a}}$. First note that both
matrices are of the form
$\left[\begin{array}{ll}I&Z\\0&I\end{array}\right]$ with $Z$
symmetric. These matrices form a commutative subgroup of the
symplectic matrices with
\[
\left[\begin{array}{cc}I&Z_a\\0&I\end{array}\right]
\left[\begin{array}{cc}I&Z_b\\0&I\end{array}\right]=
\left[\begin{array}{cc}I&Z_a+Z_b\\0&I\end{array}\right].
\]
Now, we realize $\left[\begin{array}{ll}I&Z\\0&I\end{array}\right]$
with one and two-qubit operations by first realizing the ones on
off-diagonal positions in $Z$ and then realizing the diagonal.
Entries $Z_{k,l}=Z_{l,k}=1$ are realized by operations
$e^{i(\pi/4)\tau_{\bar a}}$ with $a_k=a_l=1$ and $a_m=0$ if $m\neq k$
and $m\neq l$. These are two-qubit operations which realize the
off-diagonal part of $Z$ and as a by-product produce some diagonal.
Now this diagonal can be replaced by the diagonal of $Z$ by one-qubit
operations $e^{i(\pi/4)\tau_{\bar a}}$ with $a_k=1$ and $a_m=0$ if
$m\neq k$, which affect only the diagonal entries $Z_{k,k}$. This
completes the construction of $C$ by means of one and two-qubit
operations.

\section{Description of stabilizer states and Clifford operations
using binary quadratic forms}
\label{secdesc}

In this section we use our binary language to get further results on
stabilizer states and Clifford operations. First, we take the binary
picture of stabilizer states and their stabilizers and show how
Clifford operations act on stabilizer states in the binary picture. We
also discuss the binary equivalent of replacing one set of generators
of a stabilizer by another. Then we move to two seemingly unrelated
results. One is the expansion of a stabilizer state in the standard
basis, describing the coefficients with binary quadratic forms. The
other is a similar description of the entries of the unitary matrix of
a Clifford operation with respect to the same standard basis.

A stabilizer state $|\psi\rangle$ is the simultaneous eigenvector, with
eigenvalues $1$, of $n$ commutable Hermitian Pauli group elements
$i^{f_k}(-1)^{b_k}\tau_{s_k}$, $k=1,\ldots,n$, where
$s_k\in\mathbb{Z}_2^{2n}, k=1,\ldots,n$ are linearly independent,
$f_k,b_k\in\mathbb{Z}_2$ and $f_k=s_k^TUs_k$. The $n$ Hermitian Pauli
group elements generate a commutable subgroup of the Pauli group,
called the stabilizer ${\cal S}$ of the state.
We will assemble the vectors $s_k$ as the columns of
a matrix $S\in\mathbb{Z}_2^{2n\times n}$ and the scalars $f_k$ and
$b_k$ in vectors $f$ and $b\in\mathbb{Z}_2^{n}$. This binary
representation of stabilizer states is common in the literature of
stabilizer codes \cite{Got:97}. The fact that the Pauli group elements
are commutable is reflected by $S^TPS=0$.  One can think of $S$, $f^T$
and $b^T$ as the ``left half'' of $C$, $d^T$ and $h^T$ of
Sec.~\ref{secbin}. In the style of that section we also define
$\bar S=\left[\begin{array}{l}S\\f^T\end{array}\right]$.

If $|\psi\rangle$ is operated on by a Clifford operation $Q$,
$Q|\psi\rangle$ is a new stabilizer state whose stabilizer is given by
$Q{\cal S}Q^\dag$. As a result, the new set of generators, represented
by $\bar S'$ and $b'$ can be found by acting with $\bar C$
and $h$, representing $Q$, as in Theorem~\ref{theocb} and
Theorem~\ref{theoQ21}. One finds 
\[
 \begin{array}{ll}
   \bar S'&=\bar C\bar S\\
        b'&= b + S^T h +
             \mbox{diag}(\bar S^T\mbox{lows}(\bar C^T \bar U \bar C)\bar S)
 \end{array}
\]

The representation of ${\cal S}$ by $\bar S$ and $b$ is not unique
as they only represent one set of generators of ${\cal S}$. In the
binary language a change from one set of generators to another is
represented by an invertible linear transformation $R$ acting on the
right on $S$ and acting appropriately on $b$. By repeated application of
Lemma~\ref{lemtautau} one finds that $\bar S$ and $b$ can be transformed as
\[
 \begin{array}{ll}
   \bar S'&=\bar S R\\
        b'&=R^Tb+\mbox{diag}(R^T\mbox{lows}(\bar S^T\bar U\bar S)R)
 \end{array}
\]
Below we will refer to such a transformation as a stabilizer basis change.

Before we state the main results of this section, we show how binary
linear algebra can also be used to describe the action of a Pauli
matrix on a state, expanded in the standard basis.
\begin{equation}
\label{eqtaubin}
 \tau_a \sum_{x\in\mathbb{Z}_2^n} \psi_x |x\rangle=
        \sum_{x\in\mathbb{Z}_2^n} (-1)^{v^Tx} \psi_{x+w} |x\rangle
\end{equation}
where $a=\left[\begin{array}{c}v\\w\end{array}\right]$.
This is proved as follows. From $\sigma_x |b\rangle = |b+1\rangle$
with $b\in\mathbb{Z}_2$, we have
$\tau_{\scriptsize\left[\begin{array}{c}0\\w\end{array}\right]}\sum_x\psi_x|x\rangle=
\sum_x\psi_x|x+w\rangle=\sum_x\psi_{x+w}|x\rangle$. From $\sigma_z
|b\rangle=(-1)^b|b\rangle$, we then find Eq.~(\ref{eqtaubin}).

Now we exploit our binary language to get results about the expansion
in the standard basis of a stabilizer state as summarized in the
following theorem, for which we give a constructive proof.

\begin{theorem}
\label{theosq}
(i) If $\bar S$ and $b$ represent a stabilizer state $|\psi\rangle$ as
described above, $\bar S$ and $b$ can be transformed by an
invertible index space transformation $|x\rangle\rightarrow|T^{-1}x\rangle$
with $T\in\mathbb{Z}_2^{n\times n}$ and an invertible stabilizer basis
change $R\in\mathbb{Z}_2^{n\times n}$ into the form
\begin{equation}
\label{eqsq1}
 \begin{array}{ll}
   \bar S'&=
   \left[\begin{array}{ccc}
     T^T & 0 & 0\\
     0      & T^{-1} & 0\\
     0      & 0 & 1
   \end{array}\right] \bar S R=
   \left[\begin{array}{ccc}
      Z       & 0       & 0\\
      0       & 0       & 0\\
      0       & 0       & I_{r_c}\\
      I_{r_a} & 0       & 0\\
      0       & I_{r_b} & 0\\
      0       & 0       & 0\\
    f_a^T& 0       & 0
   \end{array}\right]\\
   b'&=\left[\begin{array}{c}
       b_{ab}\\ b_c
      \end{array}\right]
 \end{array}
\end{equation}
where $Z$ is full rank and symmetric and $f_a=\mbox{diag}(Z)$.

(ii) The state $|\psi\rangle$ can be expanded in the standard basis as
\[
\begin{array}{l}
 |\psi\rangle= (1/\sqrt{(2^{(r_a+r_b)})})\times\\
 \sum_{y\in\mathbb{Z}_2^{(r_a+r_b)}}
 (-i)^{f_a^Ty_a}(-1)^{(y_a^T\mbox{\small
 lows}(Z+f_af_a^T)y_a+b_{ab}^Ty)} |T{\scriptsize \left[\begin{array}{c}y\\ 
 b_c\end{array}\right]}\rangle 
\end{array}
\]
where $y=\left[\begin{array}{c}y_a\\y_b\end{array}\right]$ with
$y_a\in\mathbb{Z}_2^{r_a}$ and  $y_b\in\mathbb{Z}_2^{r_b}$.
\end{theorem}

In words this theorem reads as follows. If the coefficients of a
stabilizer state $|\psi\rangle$, with respect to the standard basis
$\{|x\rangle|x\in\mathbb{Z}_2^n\}$, are considered as a function of
the binary basis label $x$, this function is nonzero in an $r_a+r_b$
dimensional plane (a coset of a subspace of $\mathbb{Z}_2^n$) and the
nonzero elements are (up to a global scaling factor) equal to
$1$,$i$,$-1$ or $-i$, where the signs are given by a binary quadratic
function over the plane and $i$'s appear either in a subplane of
codimension one or nowhere (if $f_a=0$).

{\bf Proof:}
First we write $S$ as a block matrix
\[
  S=\left[\begin{array}{c}
      V\\W
    \end{array}\right]
\]
with $V,W\in\mathbb{Z}_2^{n\times n}$. Then we perform a first
stabilizer basis change $R_1$, transforming $W$ to $W^{(1)}=WR_1=[W^{(1)}_{ab}~0]$,
where $W^{(1)}_{ab}\in\mathbb{Z}_2^{n\times(r_a+r_b)}$ and
$r_a+r_b=\mbox{rank}(W)$. This is achieved by setting the last columns of
$R_1$ equal to a basis of the kernel of $W$ and choosing the other
columns as to make it invertible. As a result the columns of $W^{(1)}_{ab}$
are a basis of the range of $W$. We also write the transformation of
$V$ in block form as $V^{(1)}=VR_1=[V^{(1)}_{ab}~V^{(1)}_c]$. Because $S^{(1)}$ is full
rank, $V^{(1)}_c$ must also be full rank.

Now we perform a second stabilizer basis change
$R_2=\left[\begin{array}{ll} R_{ab,ab} & 0\\
                             R_{c,ab} & I_{r_c}
       \end{array}\right]$,
transforming $V^{(1)}=[V^{(1)}_{ab}~V^{(1)}_c] $ to $V^{(2)}=V^{(1)}R_2=[V^{(2)}_a~0~V^{(2)}_c]$, where
$V^{(2)}_a\in\mathbb{Z}_2^{n\times r_a}$ and $r_a+r_c=\mbox{rank}(V)$. This is
achieved by setting the columns $r_a+1$ till $r_a+r_b$ of $R_2$ equal
to a basis of the kernel of $V^{(1)}$ and choosing the first $r_a$ columns as
to make it invertible. (Note that the last $r_c$ columns of $R_2$ are
equal to the corresponding columns of the identity matrix and no
linear combination of them can be in the kernel of $V^{(1)}$ as
$V^{(1)}_c$ is full rank). As a result the 
columns of $[V^{(2)}_a~V^{(2)}_c]$ are a basis of the range of $V$. We also
write the transformation of $W^{(1)}$ in block form as
$W^{(2)}=W^{(1)}R_2=[W^{(2)}_a~W^{(2)}_b~0]$.

Next we perform an index space transformation
$|x\rangle\rightarrow|T^{-1}x\rangle$ with $T=[W^{(2)}_a~W^{(2)}_b~W^{(2)}_c]$
where the columns $W^{(2)}_c$ are chosen as to make $T$ invertible. As a
result $V^{(2)}$ is transformed to $V^{(3)}=T^TV^{(2)}=[V^{(3)}_a~0~V^{(3)}_c]$, $W^{(2)}$
is transformed to $W^{(3)}=T^{-1}W^{(2)}=
\left[\begin{array}{cc}
  I_{r_a+r_b} & 0\\
  0           & 0
\end{array}\right]$.  
Because $S^{(3)}=\left[\begin{array}{c}
  V^{(3)}\\W^{(3)}\end{array}\right]$
  satisfies ${S^{(3)}}^TPS^{(3)}=0$, one also finds
$V^{(3)}=\left[\begin{array}{ccc}
         Z & 0 & 0\\
         0 & 0 & 0\\
         V^{(3)}_{ca} & 0 & V^{(3)}_{cc}
      \end{array}\right]$ where $Z$ is symmetric and $V^{(3)}_{cc}$ is full rank.
A final stabilizer basis change $R_3=
  \left[\begin{array}{ccc}
    I_{r_a} & 0       & 0\\
    0       & I_{r_b} & 0\\
    {V^{(3)}_{cc}}^{-1} V^{(3)}_{ca} & 0 & {V^{(3)}_{cc}}^{-1}
  \end{array}\right]$
transforms $V^{(3)}$ to
 $V'=V^{(3)}R_3=\left[\begin{array}{ccc}
    Z & 0  & 0\\
    0 & 0  & 0\\
    0 & 0  & I_{r_c}
  \end{array}\right]$ and leaves $W^{(3)}=W'$ unchanged.
Through all the transformations we also have to keep track of $f$ and
$b$. We find $f'=\mbox{diag}(S'^TUS')=
\left[\begin{array}{c}\mbox{diag}(Z)\\0\end{array}\right]$. Setting
$R=R_1R_2R_3$ we find $\left[\begin{array}{c}
                          b_{ab}\\ b_{c}
                       \end{array}\right]=
R^Tb+\mbox{diag}(R^T\mbox{lows}(V^TW+dd^T)R)$.

We still have to prove that $Z$ is full rank. First note that
$Z={W^{(2)}_a}^TV^{(2)}_a$. From ${S^{(2)}}^TPS^{(2)}=0$ and the fact
that $[V^{(2)}_a~V^{(2)}_c]$ and $[W^{(2)}_a~W^{(2)}_b]$ are full
rank, it follows that the columns of $W^{(2)}_b$ span the orthogonal
complement of $[V^{(2)}_a~V^{(2)}_c]$ and the columns of $V^{(2)}_c$
span the orthogonal complement of $[W^{(2)}_a~W^{(2)}_b]$. Assume now
that there exists some $x\in\mathbb{Z}_2^{r_a}$ with $x\neq 0$ and
$Zx=0$, then $V^{(2)}_ax$ is orthogonal to the columns of $W^{(2)}_a$.
And $V^{(2)}_ax$ is also orthogonal to the columns of $W^{(2)}_b$.
Therefore $V^{(2)}_ax$ is a linear combination of the columns of
$V^{(2)}_c$. This is in contradiction with the fact that
$[V^{(2)}_a~V^{(2)}_c]$ is full rank.  Therefore, $Z$ is full rank.
This completes the proof of part (i).

To prove part (ii), first observe that applying $|x\rangle\rightarrow
|T^{-1}x\rangle$ to $|\psi\rangle$ simply replaces
$|T{\scriptsize\left[\begin{array}{c}y\\ b_c\end{array}\right]}\rangle$ by
$|{\scriptsize\left[\begin{array}{c}y\\ b_c\end{array}\right]}\rangle$, and
stabilizer basis transformations only change the description of a
stabilizer state but not the state itself. Therefore, we have to prove
that
\begin{equation}
\label{eqsq2}
\begin{array}{l}
 |\phi\rangle=\\
 \sum_{y\in\mathbb{Z}_2^{(r_a+r_b)}}
 (-i)^{f_a^Ty_a}(-1)^{(y_a^T\mbox{\small
 lows}(Z+f_af_a^T)y_a+b_{ab}^Ty)} |{\scriptsize\left[\begin{array}{c}y\\ 
 b_c\end{array}\right]}\rangle 
\end{array}
\end{equation}
is an eigenvector with eigenvalue one of the operators
$i^{f'_k}(-1)^{b'_k}\tau_{s'_k}$ described by $\bar S'$ and $b'$. For
$k=1,\ldots,r_a$, we have
\[
\begin{array}{ll}
 s'_k &= \left[\begin{array}{c} Ze_k\\0\\e_k\\0\end{array}\right]\\
 f'_k &= {f_a}_k = z_{k,k}\\
 b'_k &= {b_{ab}}_k
\end{array}
\]
where $e_k$ is the $k$-th column of $I_{r_a}$. With Eq.~(\ref{eqtaubin})
we find
\[
 \begin{array}{l}
  i^{f'_k}(-1)^{b'_k}\tau_{s'_k}|\phi\rangle\\
  =\sum_y[
   i^{{f_a}_k}(-1)^{{b_{ab}}_k}(-1)^{(Ze_k)^Ty_a}(-i)^{f_a^T(y_a+e_k)}\times\\
   (-1)^{((y_a+e_k)^T\mbox{\small lows}(Z+f_af_a^T)(y_a+e_k)+b_a^T(y_a+e_k)+b_b^Ty_b)}\times\\
   |{\scriptsize\left[\begin{array}{c}y\\  b_c\end{array}\right]}\rangle]\\
 = \sum_y[
   i^{{f_a}_k}(-i)^{f_a^Ty_a}(-i)^{{f_a}_k}(-1)^{f_a^Ty_a{f_a}_k}\times\\
   (-1)^{e_k^TZy_a+{b_{ab}}_k}
   (-1)^{(y_a^T\mbox{\small lows}(Z+f_af_a^T)y_a)}\times\\
   (-1)^{(e_k^T(Z+f_af_a^T)y_a +b_a^Ty_a+{b_{ab}}_k+b_b^Ty_b)}
   |{\scriptsize\left[\begin{array}{c}y\\  b_c\end{array}\right]}\rangle]\\
 = |\phi\rangle
 \end{array}\\
\]
For $k=r_a+1,\ldots,r_b$ we have
\[
\begin{array}{ll}
 s'_k &= \left[\begin{array}{c} 0\\e_k\\0\end{array}\right]\\
 f'_k &= 0 \\
 b'_k &= {b_{ab}}_k
\end{array}
\]
where now $e_k$ is the $k$-th column of $I_{(r_a+r_b)}$. With
Eq.~(\ref{eqtaubin}) we find
\[
 \begin{array}{l}
  i^{f'_k}(-1)^{b'_k}\tau_{s'_k}|\phi\rangle\\
  =\sum_y[(-1)^{{b_{ab}}_k}(-i)^{f_a^Ty_a}\times\\
          (-1)^{(y_a\mbox{\small lows}(Z+f_af_a^T)y_a+b_{ab}^T(y+e_k))}
  |{\scriptsize \left[\begin{array}{c}y\\  b_c\end{array}\right]}\rangle]\\
  =|\phi\rangle
 \end{array}
\]
For $k=r_b+1,\ldots,n$, we find with Eq.~(\ref{eqtaubin}) that
$i^{f'_k}(-1)^{b'_k}\tau_{s'_k}|x\rangle=(-1)^{x_k+b'_k}|x\rangle$. The
state $|\phi\rangle$ is clearly an eigenstate of this operator as
$x_k+b'_k=0$ for all states $|x\rangle=|{\scriptsize\left[\begin{array}{c}y\\
    b_c\end{array}\right]}\rangle$ and $k=r_b+1,\ldots,n$. 
This completes the proof. \hfill$\square$

Finally, we show how also the entries of a Clifford matrix can be
described with binary quadratic forms, by using
Theorem~\ref{theosympdec}. This leads to the following theorem for
which we give a constructive proof.

\begin{theorem}
\label{theocq}
Given a Clifford operation $Q$, represented by $\bar C$ and $h$ (or
$C$,$d$ and $h$) as in Sec.~\ref{secbin}, $Q$ can be written as
\[
\begin{array}{ll}
 Q=&  (1/\sqrt{2^r}) \sum_{x_b\in\mathbb{Z}_2^{n-r}}
    \sum_{x_r\in\mathbb{Z}_2^r}
    \sum_{x_c\in\mathbb{Z}_2^r}\\
& [(-i)^{d_{br}^Tx_{br}}(-i)^{d_{bc}^Tx_{bc}}(-1)^{(h_{bc}^Tx_{bc}
    +x_r^Tx_c)}\times\\ 
&(-1)^{x_{br}^T\mbox{\small lows}(Z_{br}+d_{br}d_{br}^T)x_{br}}\times\\
&(-1)^{x_{bc}^T\mbox{\small lows}(Z_{bc}+d_{bc}d_{bc}^T)x_{bc}}
 |T_1 x_{br} \rangle \langle T_2^{-1} x_{bc}+t|]
\end{array}
\]
where $x_{br}=\left[\begin{array}{c}x_b\\x_r\end{array}\right]$ and
$x_{bc}=\left[\begin{array}{c}x_b\\x_c\end{array}\right]$,
$T_1,T_2\in\mathbb{Z}_2^{n\times n}$ are invertible matrices,
$Z_{br},Z_{bc}\in\mathbb{Z}_2^{n\times n}$ are symmetric,
$d_{br}=\mbox{diag}(Z_{br})$, $d_{bc}=\mbox{diag}(Z_{bc})$ and
$h_{bc},t\in\mathbb{Z}_2^n$.

\end{theorem}

{\bf Proof:} 
The proof is based on the decomposition of $C$ as a product of five
matrices as in Theorem~\ref{theosympdec}. Due to the isomorphism
between the group of symplectic matrices $C$ and the extended matrices
$\bar C$ as defined in Sec.~\ref{secbin}, this decomposition
can be converted into a decomposition of $\bar C$ as follows.
\[
\begin{array}{ll}
  \bar C &= \bar C^{(1)}\bar C^{(2)}\bar C^{(3)}\bar C^{(4)}\bar C^{(5)}\\
 &= \left[\begin{array}{ccc}
     T_1^{-T} & 0   &0\\
     0        & T_1 &0\\
     0        & 0   &1
    \end{array}\right]
  \left[\begin{array}{lll}
      I_{n}    & Z_{br}   & 0\\
      0        & I_{n}    & 0\\
      0        & d_{br}^T & 1
    \end{array}\right]\times\\
  &  \left[\begin{array}{lllll}
      I_{n-r} & 0   & 0       & 0   & 0\\
      0       & 0   & 0       & I_r & 0\\
      0       & 0   & I_{n-r} & 0   & 0\\
      0       & I_r & 0       & 0   & 0\\
      0       & 0   & 0       & 0   & 1  
    \end{array}\right]
  \left[\begin{array}{lll}
      I_{n}   & Z_{bc}   & 0\\
      0       & I_{n}    & 0\\
      0       & d_{bc}^T & 1
    \end{array}\right]
     \left[\begin{array}{ccc}
     T_2^{-T} & 0   & 0\\
     0        & T_2 & 0\\
     0        & 0   & 1\\
    \end{array}\right],
\end{array}
\]
where $Z_{br}=\left[\begin{array}{cc} Z_3   & V_1\\
                                  V_1^T & Z_1
              \end{array}\right]$,
$Z_{bc}=\left[\begin{array}{cc} 0     & V_2\\
                            V_2^T & Z_2
              \end{array}\right]$,
$d_{br}=\mbox{diag}(Z_{br})$ and $d_{bc}=\mbox{diag}(Z_{bc})$.

If we define Clifford operations $Q^{(k)}$ by $\bar C^{(k)}$ and
$h^{(k)}=0$, $k=1,\ldots,5$, the operation
$Q^{(1)}Q^{(2)}Q^{(3)}Q^{(4)}Q^{(5)}$ is represented by $\bar C$ and
some vector $h'$, that can be found by repeated application of
Theorem~\ref{theoQ21}. The vector $h$ of the given Clifford operation $Q$
can then be realized by an extra operation $Q^{(6)}$ to the right with
$\bar C^{(6)}=I$ and $h^{(6)}=h+h'$. Now, $Q^{(3)}$
is a Hadamard operation on the last $r$ qubits. Because a Hadamard
operation on one qubit can be written as $H_1=(1/\sqrt{2})
\sum_{b_r,b_c\in\mathbb{Z}_2} (-1)^{b_rb_c} |b_r\rangle \langle b_c|$,
the Hadamard operation on $r$ qubits can be written as
$H_r(1/\sqrt{2^r}) \sum_{x_r,x_c\in\mathbb{Z}_2^r} (-1)^{x_r^Tx_c}
|x_r\rangle \langle x_c|$ and, including the $n-r$ qubits that are not
operated on, as
\begin{equation}
\label{eqq3}
Q^{(3)}=(1/\sqrt{2^r})
\sum_{x_b\in\mathbb{Z}_2^{n-r}}\sum_{x_r,x_c\in\mathbb{Z}_2^r} (-1)^{x_r^Tx_c} |x_{br}\rangle \langle
x_{bc}|.
\end{equation}
Considered as a matrix this is a block diagonal matrix with
$2^{n-r}$ identical $2^r\times 2^r$ blocks with entries that are $1$
or $-1$. The index $x_b$ addresses the blocks and the indices $x_c$
and $x_r$ adress the columns and rows inside the blocks. Now we will
show that the matrix $Q$ can be derived from this matrix by
multiplying on the left and the right with a diagonal matrix and a
permutation matrix representing an affine index space transformation.
First we concentrate on $Q^{(2)}$ and $Q^{(4)}$. $\bar C^{(2)}$ and
$\bar C^{(4)}$ have the form
\[
\bar{\tilde C}=\left[\begin{array}{ccc} I & \tilde Z & 0\\
                          0 & I & 0\\
                          0 & \tilde d & 1
 \end{array}\right].
\]
                       
We show that such a matrix (together with $\tilde h=0$)
represents a diagonal Clifford operation
\begin{equation}
\label{eqtq}
\tilde Q=\sum_{x\in\mathbb{Z}_2^n}
(-i)^{\tilde d^Tx} (-1)^{x^T\mbox{\small lows}(\tilde Z+\tilde
d\tilde d^T)x}|x\rangle\langle x|.
\end{equation}
This result can be derived using the decomposition in (diagonal) one
and two-qubit operations given in Sec.~\ref{sec2q}, but can more
easily be proved by showing that the Pauli group elements $\tau_{e_k}$,
with $e_k$ the $k$-th column of $I_{2n}$, are mapped to operators
represented by the columns of $\bar{\tilde C}$ under
$X\rightarrow\tilde QX\tilde Q^\dag$. Clearly, for $k=1,\ldots,n$,
$\tilde Q \tau_{e_k} \tilde Q^\dag=\tau_{e_k}\tilde Q\tilde
Q^\dag=\tau_{e_k}$ (as $\tilde Q$ and $\tau_{e_k}$ are diagonal). For
$k=n+1,\ldots,2n$ let $e_k$ again be the $k$-th column of $I_{2n}$ and
$e'_k$ the $k$-th column of $I_n$. Then we have
\[
\begin{array}{l}
 \tilde Q \tau_{e_k} \tilde Q^\dag \tau_{e_k}\\
 =\sum_{x} [(-i)^{\tilde d^Tx} (-1)^{x^T\mbox{\small lows}(\tilde Z+\tilde
d\tilde d^T)x}|x\rangle\langle x|]\times\\
 ~\sum_{x} [(+i)^{\tilde d^T(x+e'_k)}
 (-1)^{(x+e'_k)^T\mbox{\small lows}(\tilde Z+\tilde d\tilde
 d^T)(x+e'_k)}|x\rangle\langle x|]\\
=\sum_{x}[ (-i)^{\tilde d^Tx}i^{\tilde
 d^Tx}i^{\tilde d^Te'_k}(-1)^{\tilde d^Tx\tilde d^Te'_k}\times\\
 ~(-1)^{x^T(\tilde Z+\tilde d\tilde d^T)e'_k}]
=i^{\tilde d_k}\tau_{\scriptsize\left[\begin{array}{c} Ze'_k\\ 0\end{array}\right]}.
\end{array}
\]
Bringing the second $\tau_{e_k}$ from the left-hand side to the
right-hand side we finally prove Eq.~(\ref{eqtq}).

Combining Eqs.~(\ref{eqq3}) and~(\ref{eqtq}), we find
\[
\begin{array}{l}
 Q^{(2)}Q^{(3)}Q^{(4)}=\\
    (1/\sqrt{2^r})
   \sum_{x_b\in\mathbb{Z}_2^{n-r}}
   \sum_{x_r,x_c\in\mathbb{Z}_2^r}
 [(-i)^{d_{br}^Tx_{br}}(-i)^{d_{bc}^Tx_{bc}}\times\\
(-1)^{x_r^Tx_c} (-1)^{x_{br}^T\mbox{\small lows}(Z_{br}+d_{br}d_{br}^T)x_{br}}\times\\
(-1)^{x_{bc}^T\mbox{\small lows}(Z_{bc}+d_{bc}d_{bc}^T)x_{bc}}
 |x_{br} \rangle \langle x_{bc}|]
\end{array}
\]
To take into account the index space transformation $C^{(1)}$ we
simply have to replace $|x_{br}\rangle$ by $|T_1x_{br}\rangle$. For $C^{(5)}$
and $C^{(6)}$ we first define $t$ and $h_{bc}\in\mathbb{Z}_2^n$ by
writing $h^{(6)}$ as
$h^{(6)}=\left[\begin{array}{c}t\\T_2^Th_{bc}\end{array}\right]$. Then,
with Eqs.~(\ref{eqsc1}) and~(\ref{eqtaubin}) we find $\langle x_{bc}|
C^{(5)} C^{(6)}=(-1)^{h_{bc}^Tx_{bc}}\langle T_2^{-1}x_{bc}+t|$. This
completes the proof. \hfill$\square$

\section{Conclusion}
We have shown the relevance of binary linear algebra (over GF(2)) for
the theory of stabilizer states and Clifford group operations. We have
described how the Clifford group is isomorphic to a group that can be
entirely described in terms of binary linear algebra. This has led to
two schemes for the decomposition of Clifford group operations in a
product of one and two-qubit operations, and to the desription of
standard basis expansions of both stabilizer states and Clifford group
operations with binary quadratic forms.

\appendix*
\section{Proof of equation~(\ref{eqsc7})}
\label{app}

Let $e_k$ be the $k$-th column of $I_{2n}$, $k=1,\ldots,2n$. Then we
have to find the images of $\tau_{e_k}$ (Hermitian matrices) under
$X\rightarrow QXQ^\dag$ with $Q=e^{i(\pi/4)\tau_{\bar
    a}}=\frac{1}{\sqrt{2}}(I+i\tau_{\bar a})$ to yield the $k$-th
column $c_k=C e_k$ of $C$ and the $k$-th entry $h_k=e_k^T h$ of $h$.
We find
\[
 \begin{array}{l}
 i^{c_k^TUc_k}(-1)^{h_k}\tau_{c_k}\\
 ~=\frac{1}{\sqrt{2}}(I+i\tau_{\bar a}) \tau_{e_k} 
     \frac{1}{\sqrt{2}}(I-i\tau_{\bar a})\\ 
 ~= \frac{1}{2}(\tau_{e_k}+\tau_{\bar a}\tau_{e_k}\tau_{\bar a})
   +\frac{1}{2}i(\tau_{\bar a}\tau_{e_k}-\tau_{e_k}\tau_{\bar a})\\
 ~= \frac{1}{2} (1+(-1)^{e_k^TPa})\tau_{e_k}
   +\frac{1}{2}i(1-(-1)^{e_k^TPa})\tau_{\bar a}\tau_{e_k},
 \end{array}
\]
where in the last step we used $\tau_{\bar a}^2=I$ and
$\tau_a\tau_b=(-1)^{b^TPa}\tau_b\tau_a$ as follows from
Lemma~\ref{lemtautau}. When $e_k^TPa=0$ we find $c_k=e_k$ and $h_k=0$. When
$e_k^TPa=1$ we find
\[
 \begin{array}{ll}
 i^{c_k^TUc_k}(-1)^{h_k}\tau_{c_k} 
   &= i \tau_{\bar a}\tau_{e_k}\\
   &= i i^{a^TUa} (-1)^{e_k^TUa} \tau_{a+e_k},
 \end{array}
\]
From this formula it can be read that $c_k=a+e_k$.
With $ii^{a^TUa}=i^{a^TUa+1}(-1)^{a^TUa}$ (with the addition in the
exponents modulo $2$) and
$(a+e_k)^TU(a+e_k)=a^TUa+e_k^TPa+e_k^TUe_k=a^TUa+1$, we also find that
$h_k=a^TUa+e_k^TUa$.

Combining the two cases $e_k^TPa=0$ and $e_k^TPa=1$ we find
$c_k=e_k+a(e_k^TPa)=(I+aa^TP)e_k$, yielding $C=(I+aa^TP)$. For $h$ we
find $h_k=(e_k^TPa)(a^TUa+e_k^TUa)$. With $(e_k^TPa)(e_k^TUa)=e_k^TUa$
this reduces to $h_k=e_k^T(Paa^TUa+Ua)$ and $h=(I+aa^TP)^TUa$. This
completes the proof. \hfill$\square$

\begin{acknowledgments}
  We thank Frank Verstraete for useful discussions.  Our research is
  supported by grants from several funding agencies and sources:
  Research Council KULeuven: Concerted Research Action GOA-Mefisto~666
  (Mathematical Engineering); Flemish Government: Fund for Scientific
  Research Flanders: several PhD/postdoc grants, projects G.0240.99
  (multilinear algebra), G.0120.03 (QIT), research communities ICCoS,
  ANMMM; Belgian Federal Government: DWTC (IUAP IV-02 (1996-2001) and
  IUAP V-22 (2002-2006): Dynamical Systems and Control: Computation,
  Identification \& Modelling
\end{acknowledgments}


\end{document}